\documentstyle[psfig,sprocl,here]{article}

\newcommand{\be}{\begin{equation}}
\newcommand{\ee}{\end{equation}}
\newcommand{\bea}{\begin{eqnarray}}
\newcommand{\eea}{\end{eqnarray}}
\newcommand{\bean}{\begin{eqnarray*}}
\newcommand{\eean}{\end{eqnarray*}}

\newcommand{\ba}{\begin{array}}
\newcommand{\ea}{\end{array}}

\begin{document}

\title{POLARIZED STRUCTURE FUNCTIONS IN QCD
  \footnote{Invited talk presented at the Workshop on Lepton Scattering,
        Hadrons and QCD, Adelaide, March 26 -- April 6, 2001.}
           }

\author{J. KODAIRA}

\address{Department of Physics, Hiroshima University, \\
Higashi-Hiroshima 739-8526, JAPAN} 


\maketitle
\abstracts{
We review the nucleon's polarized structure functions 
from the viewpoint of gauge invariant, nonlocal light-cone operators in QCD.
We discuss a systematic treatment of
the polarized structure functions and the corresponding
parton distribution functions. 
We also address a question of what information on the structure
of Nature will be obtained from the future polarized experiments.
From this point of view, we will discuss the $W^{\pm}\gamma$
production at RHIC polarized experiment.
}

\section{Introduction}

In the last ten years, great progress has been made
both theoretically and experimentally in hadron spin physics.
Furthermore, in conjunction with 
new projects like the \lq\lq RHIC spin project\rq\rq, \lq\lq polarized
HERA\rq\rq , etc.,  we are now
in a position to obtain more information on the 
spin structure of nucleons.
The spin dependent quantity is, in general, very sensitive to the
structure of interactions among various particles. 
Therefore, we will be able to study the detailed structure of hadrons
based on QCD.
However the purpose of new experiments should not be limited to only
the check of QCD.
We also hope that we can find some clue to new physics beyond the
standard model.

In this talk, we first review a systematic treatment of polarized
structure functions and summarize the recent theoretical progress on
the QCD evolutions~\cite{kodatana}.
Secondly, we address a question of what we will be able to learn
from the future polarized experiments by considering the phenomena
called Radiation Zeros (RAZ)~\cite{BM} as an example.
We reanalyze this phenomena at the realistic RHIC 
polarized collider.
We will point out that the polarization of colliding protons will
emphasize the RAZ phenomena in the cross section and a \lq\lq moderate
energy\rq\rq\ machine is better than the extremely high energy
machines to find this phenomena.

\section{Classification of structure functions}

To describe a variety of high-energy processes to which
the factorization theorem can be applied,
it is desirable to have a formulation based on a universal language in QCD.
The traditional approach relies on the operator product expansion (OPE), 
but it can be applied only to a limited class of processes.
This calls for an approach based on the factorization
as a generalization of the OPE.
As a result of factorization, the structure functions (cross section)
are given as convolution of the short- and long- distance parts.
The former contains all the dependence on the hard scale and the
latter is controlled by the nonperturbative dynamics of QCD.

For the long-distance part, the parton distribution functions
(PDFs) are introduced and it is now standard to define them
as the nucleon's matrix elements of nonlocal light-cone operators
in QCD~\cite{CS}.
The momentum of nucleon (mass $M$) $P$ will be written in terms of 
two auxiliary light-like vectors $p^{\mu}$ and $w^{\mu}$
$(p^2 = w^2 = 0\ , \ p\cdot w=0)$,
$P^{\mu} = p^{\mu} + \frac{M^2}2 w^{\mu}$.
The (quark) PDF is defined by
the Fourier transform of the nucleon's matrix element of 
the nonlocal light-cone quark operator as,
\be
 \int_{-\infty}^{+\infty} \frac{d\lambda}{2\pi} e^{i\lambda x} 
        \langle P S|
       \bar{\psi}(0) [0,\lambda w] \Gamma \psi(\lambda w) |P S \rangle 
    \  . \label{eq:gam}
\ee
Here $|P S \rangle$ is the nucleon state with spin $S$ which is
decomposed into the longitudinal and transverse parts with respect
to $p^{\mu} (w^{\mu})$ direction,
$S^{\mu} = S_{\parallel}^{\mu} + S_{\perp}^{\mu}$.
The variable $x$ can be interpreted as the momentum fraction of
parton (Bjorken $x$) since $\lambda x = (xP)\cdot (\lambda w)$. 
$\Gamma$ is a generic Dirac matrix and
the link operator $[0, \lambda w]$ which makes the operators gauge
invariant is given by,
\[    [y,z]= {\rm P}\! \exp \left(ig \!\! \int_0^1 \!\!\! dt
      (y-z)_\mu A^\mu(ty+(1-t)z) \right)\ . \]
Possible choices for $\Gamma$ and spin states which depend
on the processes considered in Eq.(\ref{eq:gam}) lead to
various PDFs.  

\subsection{Quark distributions}

Let us first consider the quark distributions.
An important observation made in Ref.[4]
is that one can generate all quark distribution functions 
up to twist-4 by substituting all the possible $\Gamma$.
By decomposing Eq.(\ref{eq:gam}) into independent tensor 
structures, one finds nine independent quark distribution functions
associating with each tensor structure~\cite{kodatana}.
Their spin, twist and chiral classifications 
are listed in Table 1.
The distributions in the first row are spin-independent, 
while those in the second and third rows correspond to the 
longitudinally ($S_{\parallel}$) and transversely ($S_{\perp}$)
polarized nucleons.
Each column refers to the twist.
The distributions marked with ``$\star$'' 
are referred to as chiral-odd, because they correspond to
chirality-violating Dirac matrix structures
$\Gamma = \{\sigma_{\mu \nu} i \gamma_{5},\, 1\}$.
The other distributions are chiral-even, because of the
chirality-conserving structures
$\Gamma = \{\gamma_{\mu},\, \gamma_{\mu}\gamma_{5}\}$.
In the massless quark limit, chirality
is conserved 
\newlength{\minitwocolumn}
\setlength{\minitwocolumn}{0.5\textwidth}
\addtolength{\minitwocolumn}{-0.5\columnsep}
\begin{minipage}[t]{\minitwocolumn}
\vspace{1ex}
{\small
\begin{center}
{\footnotesize
Table 1:\ Spin,twist and chiral classification of
       quark distributions.}

\vspace{1ex}
\begin{tabular}{|c|ccc|}\hline
Twist    & 2  & 3 & 4  \\ \hline
$S$ ave.& $q(x)$ & $e(x)^{\star}$ & $f_{4}(x)$ \\
$S_{\parallel}$ & $\Delta q(x)$ & $h_{L}(x)^{\star}$ 
& $g_{3}(x)$ \\
$S_{\perp}$ &  $\delta q(x)^{\star}$ & 
$g_{T}(x) $ &  $h_{3}(x)^{\star}$\\
\hline
\end{tabular}
\end{center}
}
\end{minipage}
\hspace{\columnsep}
\begin{minipage}[t]{0.95\minitwocolumn}
through the propagation of a quark.
This means that in the DIS one can measure only the chiral-even 
distributions up to tiny quark mass corrections
because the perturbative quark-gluon 
and quark-photon couplings conserve the chiralities.
On the other hand, in the Drell-Yan and
\end{minipage}

\vspace{0.3ex}
\noindent
certain other processes, both chiral-odd and chiral-even distribution
functions can be measured~\cite{RS} because the chiralities of the 
quark lines originating in a single nucleon 
are uncorrelated.

\subsection{Gluon distributions}

The gluon distribution functions can be defined in the similar way as
for the quark distributions.
The gauge-invariant definition of the gluon distribution functions
is provided by~\cite{CS,mano},
\[  \frac{2}{x} \int \frac{d\lambda}{2\pi} e^{i\lambda x}
   \langle PS| w_{\alpha} G^{\alpha \mu}(0) [0, \lambda w]
  w_{\beta} G^{\beta \nu}(\lambda w) |PS \rangle \ . \]
Corresponding to independent tensor structures, we have four gluon
distribu-
\begin{minipage}[t]{\minitwocolumn}
\vspace{1ex}
{\small
\begin{center}
{\footnotesize
Table 2:\ Spin and twist classification of
       gluon distribution functions.}

\vspace{1ex}
\begin{tabular}{|c|ccc|}\hline
Twist    & 2 & 3 & 4 \\ \hline
$S$ ave.& ${\cal G}(x)$ &  & ${\cal G}_{4}(x)$ \\
$S_{\parallel}$ &  $\Delta {\cal G}(x)$  & &  \\
$S_{\perp}$ &  & ${\cal G}_{3T}(x)$ & \\ \hline
\end{tabular}
\end{center}
}
\end{minipage}
\hspace{\columnsep}
\begin{minipage}[t]{0.95\minitwocolumn}
tion functions summarized in Table 2.
The gluon distributions mix through renormalization
with the flavor singlet chiral-even quark distributions.
On the other hand, there exists no gluon distributions
that mix with the chiral-odd quark distributions.
\end{minipage}

\subsection{Twist-3 three particle distributions}

Coherent many-particle contents of the nucleon are described by
multi PDFs.
In this talk, we mention only the twist-3 quark-gluon
correlation functions,
\[ \int \frac{d\lambda}{2\pi}\frac{d\zeta}{2\pi}
   \,e^{i\lambda x + i\zeta (x'-x)}
  \langle PS|\bar{\psi}(0) \Gamma [0,\zeta w] g\, G^{\mu\nu}(\zeta w)
     [\zeta w,\lambda w] \psi(\lambda w)|PS\rangle .\]
Similarly to the quark distributions, one can define the  
multiparton distributions by considering possible Dirac matrices for
$\Gamma$ which are listed in Table 3.
\begin{minipage}[t]{\minitwocolumn}
\vspace{1ex}
{\small
\begin{center}
{\footnotesize
Table 3:\ Spin and chiral classification of
   quark-gluon correlations at twist-3.}

\vspace{1ex}
\begin{tabular}{|c|c|}\hline
$S$ ave. & $\Phi (x, x')^{\star}$\\
$S_{\parallel}$  &
$\widetilde{\Phi} (x, x')^{\star}$\\  
$S_{\perp}$  &  $\Psi (x, x') \  , \ 
 \widetilde{\Psi} (x, x')$ \\ \hline
\end{tabular}
\end{center}
}
\end{minipage}
\hspace{\columnsep}
\begin{minipage}[t]{0.95\minitwocolumn}
The treatment here can be extended to the case of 
three-gluon correlation functions, which are relevant to 
the (singlet) quark distribution $g_{T}(x)$ and the gluon distribution
${\cal G}_{3T}(x)$.
The details have been discussed in Refs.[6,7].
\end{minipage}

\section{QCD evolution}

In the case of the lowest twist, their QCD evolution
can be easily estimated using the standard techniques.
The calculation for the higher twist terms is, however,
generally very complicated due to the presence of multiparton
distributions.
Although it is possible to generalize the DGLAP approach
to the three-body case,
one is forced to use some particular techniques, e.g.
the use of the light-like axial gauge etc.~\cite{BKL,DMu}
On the other hand, it has been known that
one can construct a set of \lq\lq local composite operators\rq\rq\ 
from the nonlocal one by taking the moment of PDFs.
Therefore we can take an approach based on these local composite 
operators~\cite{KODS}.
The advantage of this approach is that all the relevant steps
can be worked out based on the standard and familiar field theory
techniques in any gauge.

The common feature for the higher twist operators is that
the number of a set of local operators
with the same quantum numbers increases with
spin (moment) and these operators are 
related through the QCD equation of motion~\cite{P80,SV}.
However, once a complete set of operators is identified,
the anomalous dimensions which derive
QCD evolution is easily calculated following
the theorem~\cite{JL76} explained now.
This theorem tells us that in non-abelian gauge theory,
three kinds of operators will mix under 
renormalization of gauge invariant operators.
(I)the gauge invariant operator itself.
(II)the BRST invariant operators.
(III)the operators which are proportional to the equation of 
motion (EOM).
The EOM operators involve both BRST ``invariant'' and
``variant'' ones~\cite{KODS,KOD5}.
Although the physical matrix elements of the BRST
invariant and EOM operators
vanish~\cite{kodatana}, it is necessary to consider these operators to
complete the renormalization.
At the lowest twist level, these complexities do not come into play
because there exists neither EOM nor BRST invariant
operator of twist-2. 
The EOM as well as the BRST invariant operators always have smaller
spin by at least one unit than the possible highest spin operators
of the same dimension.

The explicit calculations at the one-loop level
appear in Refs.[10,8,15] for $g_T$ and in Ref.[16]
for $h_L$.
Much more recent progress on the twist-3 parton densities can be
found in Ref.[17].
 
\section{$W^{\pm}\gamma$ production at RHIC}

The polarized PDFs will be measured precisely 
by the future polarized experiments and we will be able to check
if the QCD predictions are consistent or not.
However, we believe that the polarized experiments
will provide us with a great store of knowledge on the
structure of all interactions. 
From this point of view, the final part of this talk is
devoted to a discussion on the
radiative weak boson production $pp \to W^{\pm}\gamma$
at RHIC polarized experiment.
We point out two reasons why we reanalyze this process.
(1) Due to the $V-A$ structure of the $W$ boson interaction,
only the initial quarks which have definite helicities can
participate in the process.
Therefore, an experiment with the polarized beams will be more
efficient to study this process~\cite{WSCY} and an information
on $\Delta q(x)$ is relevant.
(2) Many works so far assumed rather high energy collisions~\cite{DS}.
However, a realistic experiment with the polarized beams
becomes available at RHIC whose center of mass
energy is around $\sqrt{s} \sim 500\ {\rm GeV}$.
   
\subsection{RAZ at the partonic level}

Radiative weak boson ($W \gamma$) production in hadronic collisions
has been the subject of
much theoretical interests since this process contains the gauge
boson trilinear coupling and develops the so-called radiation
zero (RAZ)~\cite{BM}.
The RAZ is a typical example which is sensitive
to the structure of electroweak interaction.

The RAZ is a phenomena that the cross section (amplitude)
for some process develops zero in the some point of the phase space.  
To understand this phenomena in the simplest way, although
not rigorous, let us use the soft photon approximation.
Since the soft photon factorizes as the eikonal factor,
the amplitude which contains one photon can be written as,
$M_{\gamma} \simeq e\, J \cdot \epsilon (k) M$ where
$J^{\mu} = \sum_i \, Q_i \, \eta_{\,i} \, \frac{p_i^{\mu}}{p_i \cdot k}$
and $\epsilon^{\mu} (k)$ is the polarization vector of photon.
$Q_i$ is the charge of $i$-the particle and
$\eta_{\, i}= + (-)$ for the incoming (outgoing) particle.
The sum is taken over all external particles.
If $\frac{Q_i}{p_i \cdot k} =$  const. for all $i$,
the energy-momentum conservation $\sum_i \eta_{i}\, p_i^{\mu} = 0$
implies $J^{\mu} = M_{\gamma} = 0$.
The well-known example is the process,
$u(p_1 ) + \bar{d} (p_2 ) \to W^+ (p_3 ) + \gamma (k)$.
In this case, the identity
$\frac{2}{3} \frac{1}{p_1 \cdot k} = \frac{1}{3} \frac{1}{p_2 \cdot k}
           = \frac{1}{p_3 \cdot k}$
is satisfied at $\cos \theta_{\gamma} = - \frac{1}{3}$.

\subsection{Hadronic cross section}

\begin{figure}[H]
\begin{center}
\begin{tabular}{cccc}
\leavevmode\psfig{file=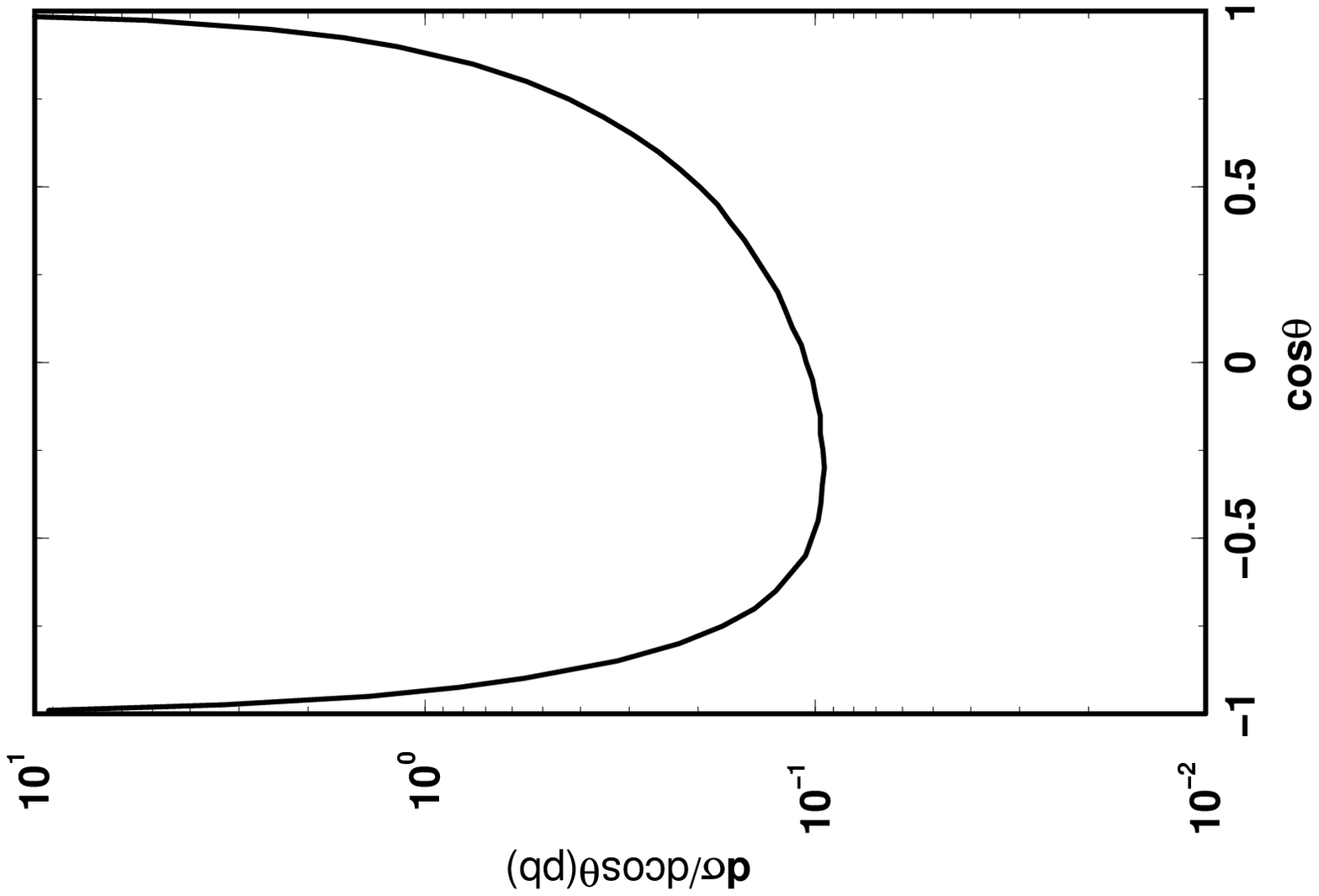,width=2.3cm,angle=-90}  &
\leavevmode\psfig{file=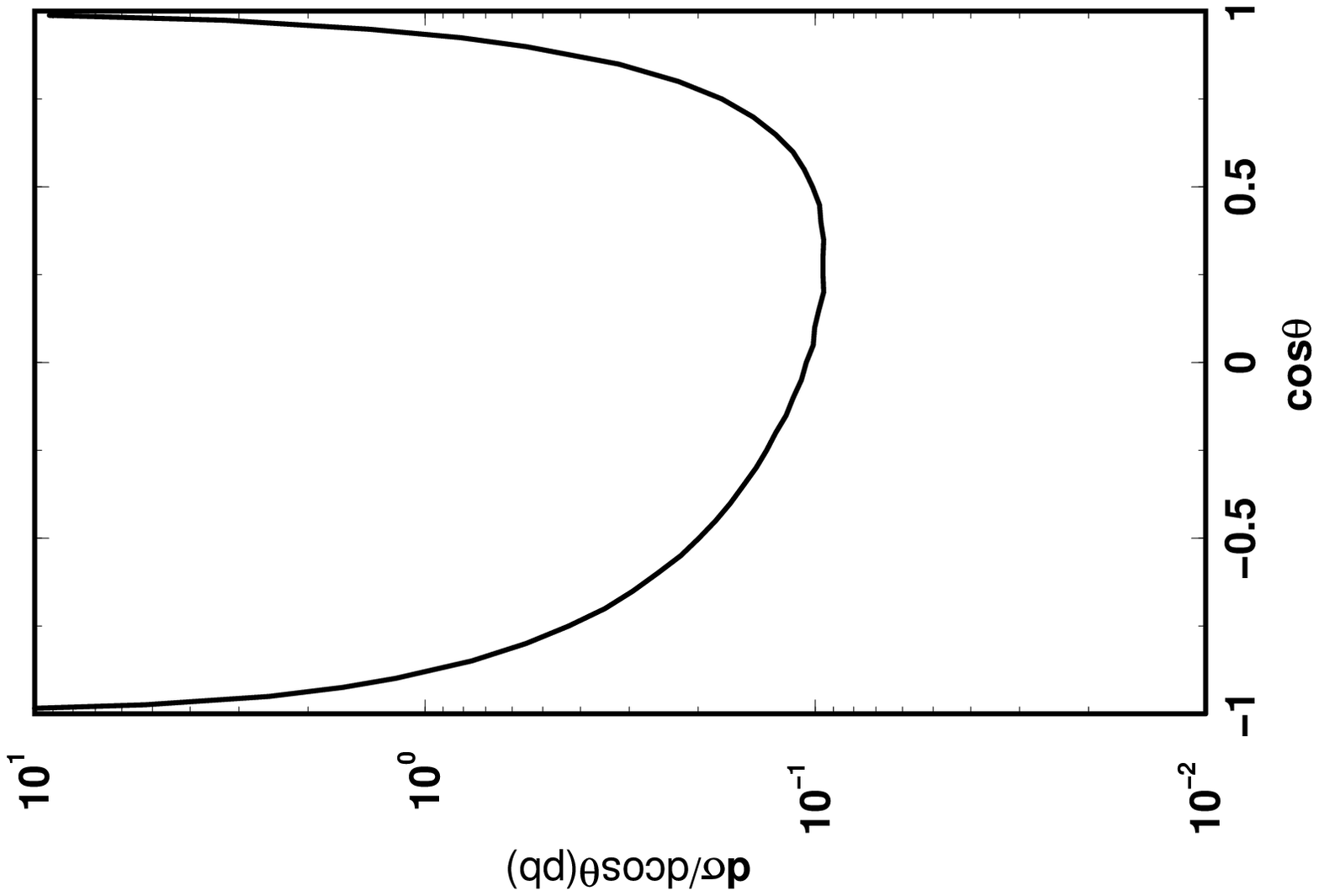,width=2.3cm,angle=-90}  &
\leavevmode\psfig{file=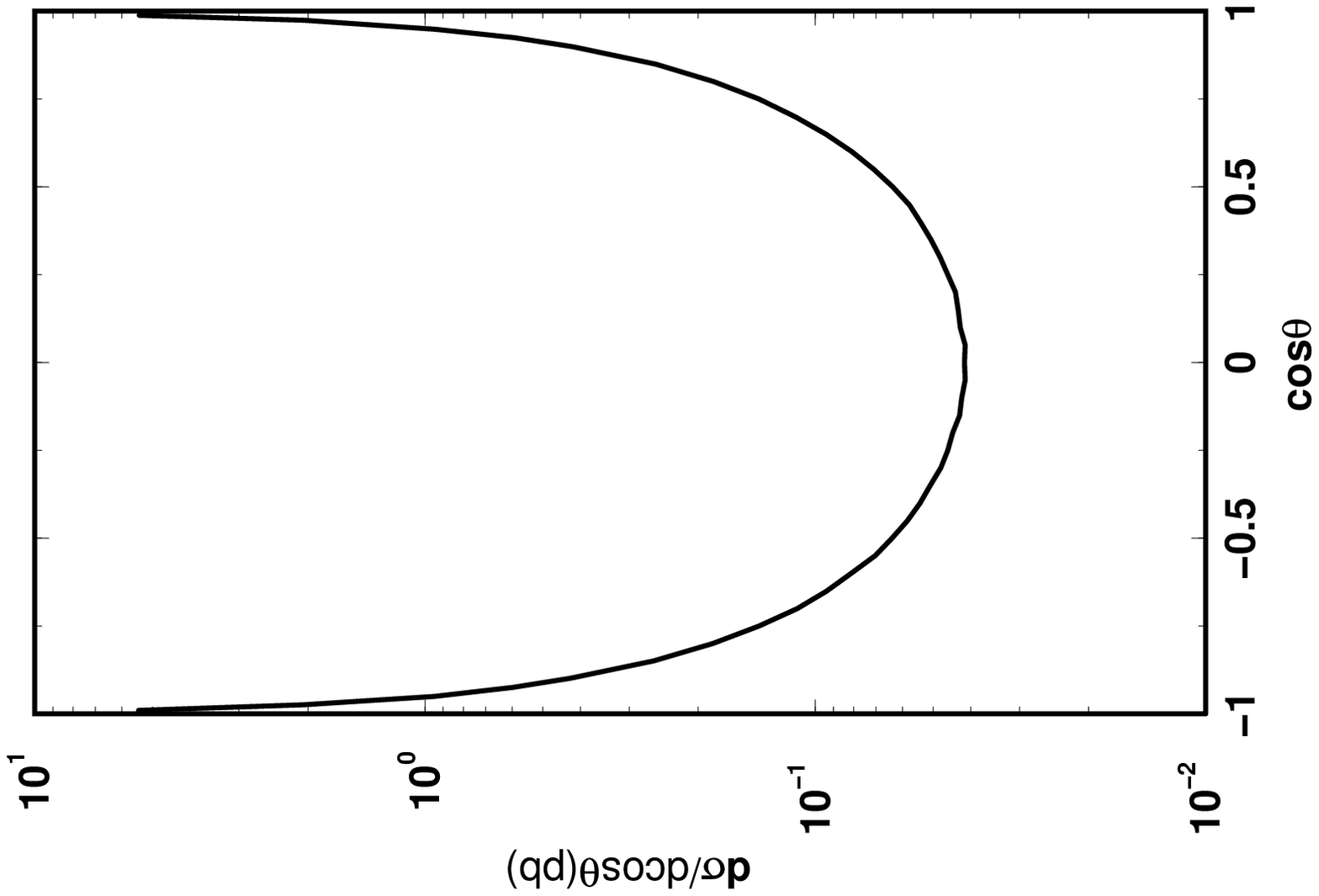,width=2.3cm,angle=-90} & 
\leavevmode\psfig{file=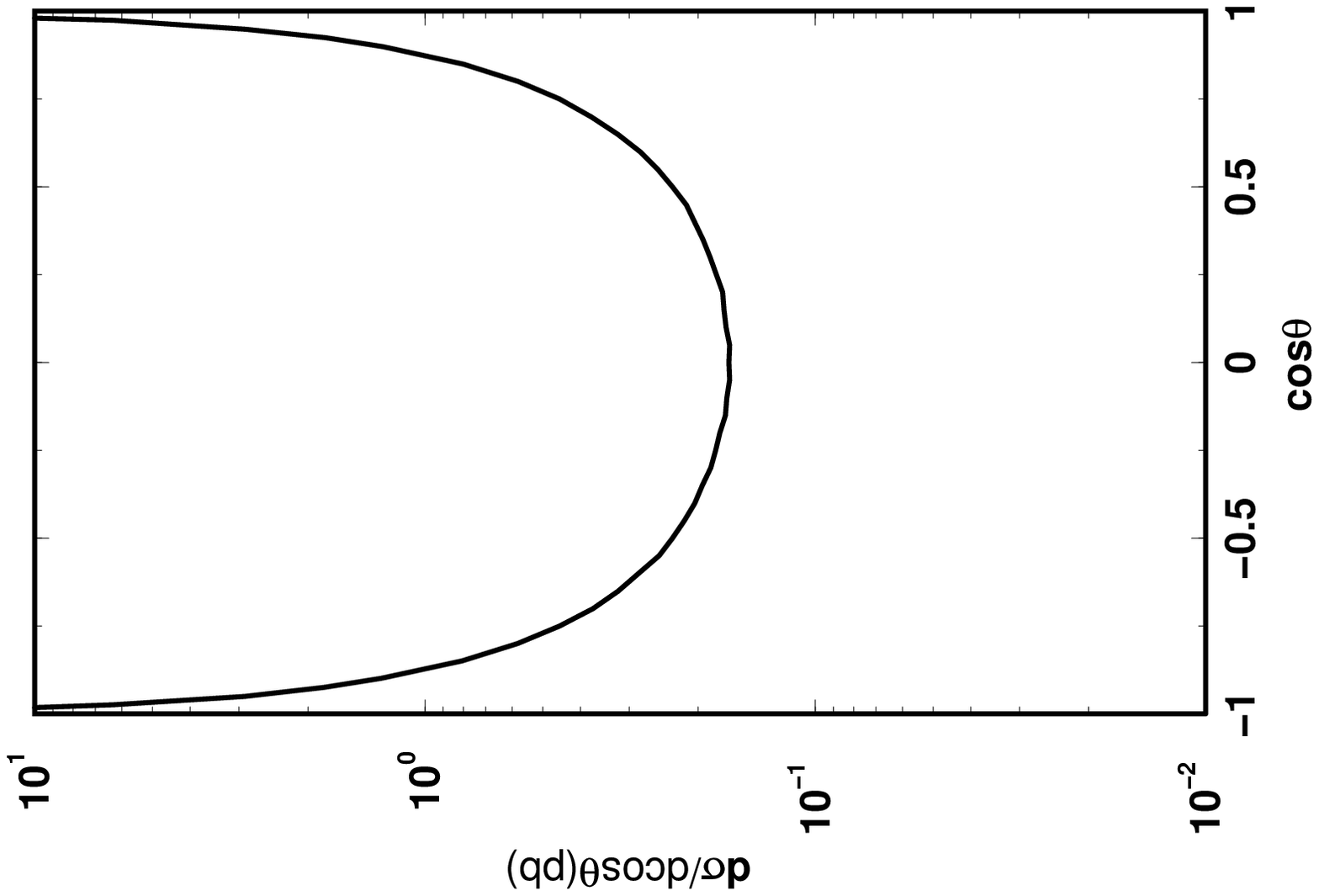,width=2.3cm,angle=-90} \\
{\footnotesize \ (a1)} &   {\footnotesize \ (a2)}
    & {\footnotesize \ (b1)} & {\footnotesize \ (b2)}
\end{tabular}
\vspace{-0.3cm}
\caption{Distribution of $\cos \theta_{\gamma}$ in the laboratory
frame for the reaction $pp \to W^+ \gamma + X$.} 
\label{fig1} 
\end{center}
\end{figure}

To obtain a realistic (hadronic) cross section,
we must convolute the partonic cross section
with the PDFs~\cite{HR}.
Fig.\ref{fig1} shows the cross section in $Pb$ for the 
various proton's spin configurations at $\sqrt{s} = 500 {\rm Gev}$.
Denoting the helicity of initial protons by $p(\pm)$,
Figs.(1.a1),(1.a2),(1.b1) and (1.b2) correspond to $p(-) p(+)$,
$p(+) p(-)$, $p(+) p(+)$ and $p(-) p(-)$ respectively.
We have chosen the minimum cut-off energy for the photon
to be $5$ Gev.

\vspace{-1.5ex}
One can see a rather clear dip in the cross section when
the initial proton's helicities are parallel each other.
This is because the helicity distributions of quarks
depend on the spin of the parent protons.
Note that if the proton contains the equal
parton densities of both helicity states as for the unpolarized case,
the convolution completely smears out the RAZ. 
How about is the energy dependence of this smearing effect
coming from the convolution?
It is easily supposed that the dip will be more smeared as
the energy becomes higher.
It is because the small $x$ partons start
to participate in the process
at higher energy and those sea partons carry less
information of the parent proton's spin.
Namely the contribution from the small $x$ partons is almost the same
for the polarized and unpolarized protons.

We also define an asymmetry by,
\[  A = \frac{ d \sigma \, ( p(-) p(+) \to W^+ \gamma )
         - d \sigma \, ( p(+) p(-) \to W^+ \gamma )}
       {d \sigma \, ( p(-) p(+) \to W^+ \gamma )
       + d \sigma \, ( p(+) p(-) \to W^+ \gamma )}
       \ ,\]
and plot it in Fig.2.
This asymmetry amounts to 40 \%.

\begin{minipage}[t]{0.85\minitwocolumn}
\vspace{-3ex}
\begin{figure}[H]
\begin{center}
\leavevmode\psfig{file=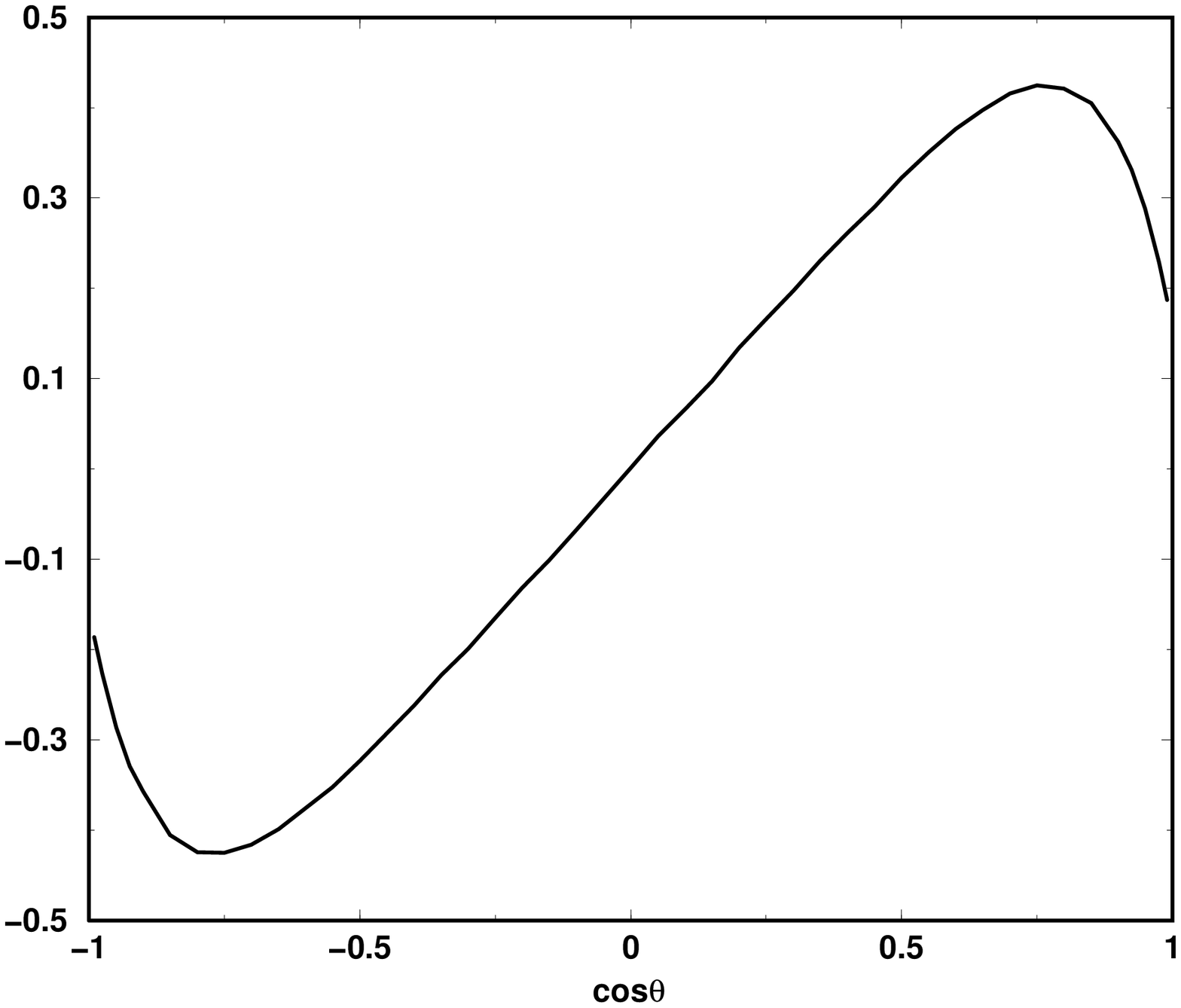,width=4cm,angle=-90}

\vspace{0.3ex}
{\footnotesize
Figure 2;\ Asymmetry w.r.t. $\cos \theta_{\gamma}$.}
\end{center}
\end{figure}
\end{minipage}
\hspace{\columnsep}
\begin{minipage}[t]{\minitwocolumn}
\hspace{1em}
A comment is in order concerning the higher order 
corrections.
Since the RAZ occurs only at the partonic tree level,
the higher order (QCD) corrections might give an important
effects and it is possible that RAZ is completely smeared out
in the physical hadronic cross sections.
There has been much effort to estimate higher order
corrections in the standard mod-
\end{minipage}

\noindent
el and phenomenological analyses~\cite{BHZO}.
The QCD radiative corrections can be classified into three effects:
(1) virtual corrections and soft gluon emissions (2)
hard gluon emissions (3) gluon initiated processes $q_1 g \to V \gamma q_2$
and $g \bar{q}_2 \to V \gamma \bar{q}_1$.
Among these effects, the first one leaves the RAZ intact.
Its main effect is an almost constant $K$ factor.
The second and third effects, on the other hand, will completely wash
out the RAZ phenomena.
These effects, however, have been known to be important at high energies.
It is, therefore, expected that in the RHIC energy region, 
it is sufficient to include only the first effect
which will not change the shape of the
cross section from the tree level one 
except for some multiplicative enhancement.
In particular, the asymmetry, Fig.2, remains the same.
 
\section{Conclusions}

We have surveyed the polarized 
structure functions from a viewpoint of gauge invariant nonlocal 
operators.
We explained an approach to obtain the QCD evolution which
preserves maximal (BRST and Lorentz) symmetries
of the theory at every step of investigation.
We also presented the radiative weak boson production
at RHIC which strongly depends on the helicity
distribution of quarks inside the proton.
We have pointed out that the experiments in the RHIC energy region
will be very efficient to study this process.

We hope that various kinds of new experiments
and theoretical investigations will be able to
clarify not only perturbative and nonperturbative
aspects of QCD but also the full structure of all interactions
in Nature.

\section*{Acknowledgments}
The author would like to thank the organizers
for their hospitality and inviting him to this stimulating workshop.
He also thanks H. Kawamura, K. Morii and K. Tanaka
for collaborations.
This work is supported in part by the Monbu-kagaku-sho Grant-in-Aid
for Scientific Research No. C-13640289.

\end{document}